\documentclass{PoS}

\def\beq{\begin{equation}}
\def\enq{\end{equation}}

\def\Tr{{\rm Tr}\,}
\usepackage{lineno}
\usepackage{braket}
\usepackage{amsmath}
\title{Tensor RG calculations and quantum simulations near criticality}
\ShortTitle{TRG and quantum simulations}

\author{\speaker{ Y. Meurice}$^a$, A. Bazavov$^{a,b,d}$, Shan-Wen Tsai$^a$,
J. Unmuth-Yockey$^b$, Li-Ping Yang$^c$, Jin Zhang$^a$\\
\llap{$^a$}Department of Physics and Astronomy, University of Iowa, Iowa City, IA 52242, USA\\
\llap{$^b$}Department of Physics and Astronomy, University of California, Riverside, CA 92521, USA\\
\llap{$^c$}Department of Physics,Chongqing University, Chongqing 401331, China \\
\llap{$^d$}Department of Physics and Astronomy, Michigan State University, East Lansing, Michigan 48824, USA\\
      E-mail: \email{yannick-meurice@uiowa.edu}}

\abstract{ We discuss the reformulation of the O(2) model with a chemical potential and the Abelian Higgs model on a 1+1 dimensional space-time lattice using the Tensor Renormalization Group (TRG) method. The TRG allows exact blocking and connects smoothly the classical Lagrangian approach to the quantum Hamiltonian approach. We calculate the entanglement entropy in the superfluid phase of the O(2) model and show that it approximately obeys the logarithmic Calabrese-Cardy scaling obtained from Conformal Field Theory (CFT). We calculate the Polyakov loop in the Abelian Higgs model and discuss the possibility of a deconfinement transition at finite volume. We propose Bose-Hubbard Hamiltonians implementable on optical lattices as quantum simulators for CFT models. 
}

\FullConference{34th annual International Symposium on Lattice Field Theory\\
		24-30 July 2016\\
		University of Southampton, UK}

\begin{document}

\section{Introduction}
Recent  calculation using the Tensor Renormalization Group (TRG) method \cite{PhysRevB.86.045139,prb87,efratirmp,Exactblocking13prd,prd89,Shimizu:2014uva,Takeda:2014vwa,Kawauchi:2015heu} have shown that blocking methods can 
complement sampling methods for models studied in lattice field theory. The ultimate goal of the method is to approach 
the continuum limit of lattice models in terms of fixed points of the TRG transformation \cite{prb87,efratirmp}. A first step in this direction is to control 
the finite dimensional approximations. 

The $O(2)$ model with a chemical potential in 1+1 dimensions has a complex action and a sign problem if simulated with MC methods,  but can be handled with a worm algorithm  \cite{Banerjee:2010kc} and the TRG method \cite{prd89,PhysRevA.90.063603}. 
For sufficiently large $\beta$ or $\mu$, the model has a superfluid phase where
we expect to have a conformal field theory (CFT) with central charge $c=1$. This idea can be tested by calculating the 
von Neumann and R\'enyi entanglement entropy (EE) and comparing their finite size scaling with the  CFT 
Calabrese-Cardy scaling  \cite{Calabrese:2004eu} which predicts a logarithmic growth with the size of the 
system and a coefficient which is the central charge divided by an integer depending on the type of entropy and the 
boundary conditions. This scaling appears very clearly for the von Neumann EE with periodic boundary conditions, but 
it was realized after the conference that for the second order R\'enyi entropy, $S_2$, with open boundary conditions, 
large subleading corrections appear and need to be taken into account.  Fortunately, they can also be understood in the context of CFT \cite{parity} 
and allowed us to check the validity of the expectation that $c=1$ \cite{preprint,draft}. 

By gauging the $O(2)$ model, we obtain the 
Abelian Higgs model. By keeping the gauge coupling small enough when $N_s^2$ increases, with $N_s$ the spatial number of sites, we can use the gauge field to 
probe  the $O(2)$ model. Below, we discuss the data collapse for the Polyakov loop calculated in the scaling limit where $g^2$ 
decreases like $1/N_s^2$. We show that this observable singles out the transition to topological order more sharply than $S_2$.

Recently, it has been possible to measure $S_2$ in quantum simulations for the Bose-Hubbard model using optical lattices \cite{Islam2015}  with a beamsplitter operation proposed in Ref.  \cite{PhysRevLett.109.020505}.  This experimental setup can 
quantum simulate the $O(2)$ model with a chemical potential  \cite{PhysRevA.90.063603} using only one species of bosonic atom provided that the chemical potential is large enough. After the conference, we 
developed more specific experimental methods to measure the central charge using existing optical lattice experiments 
\cite{preprint}. This opens the possibility of exploring CFT using one-dimensional optical lattices.

\section{The Tensor Renormalization Group (TRG) method}
The TRG method provides {\it exact} blocking formulas for spin and gauge models \cite{Exactblocking13prd}. 
 The blocking separates the degrees of  freedom inside the block (integrated over), from those kept to communicate with the neighboring blocks. The only approximation is the truncation in the number of ``states" kept.  It 
applies to many lattice models: the Ising model \cite{PhysRevB.86.045139,prb87,efratirmp}, the $O(2)$ model \cite{Exactblocking13prd,sb}, the $O(3)$ model \cite{Exactblocking13prd,af} ,  the $SU(2)$ principal chiral model, Abelian and $SU(2)$ gauge theories  \cite{Exactblocking13prd}, the Schwinger model \cite{Shimizu:2014uva,Saito:2014bda}, Gross-Neveu model \cite{Kawauchi:2015heu}, 
and CP(N-1) models \cite{Takeda:2014vwa}. More recent progress has been reviewed in the plenary talk by Shinji Takeda \cite{st}. 

The TRG method relies on  algebraic manipulations of positive matrices and seems free of sign problems. 
It can handle problems with complex temperature \cite{prd89} and real chemical potential  \cite{PhysRevA.90.063603} .
It has been checked \cite{PhysRevE.93.012138} with a worm algorithm \cite{Banerjee:2010kc}. 
It has been used to study fixed points \cite{prb87,efratirmp}, however the effects of truncations are not well-understood in this context. 
The TRG approach connects easily to the Hamiltonian picture and provides spectra which can be used for real time calculations. It has been 
used to design quantum simulators for 
the $O(2)$ model  \cite{PhysRevA.90.063603}  and the Abelian Higgs model  \cite{PhysRevD.92.076003}.

\section{The R\'enyi entanglement entropy} 
\begin{figure}[h]
\begin{center}
\includegraphics[width=0.4\textwidth]{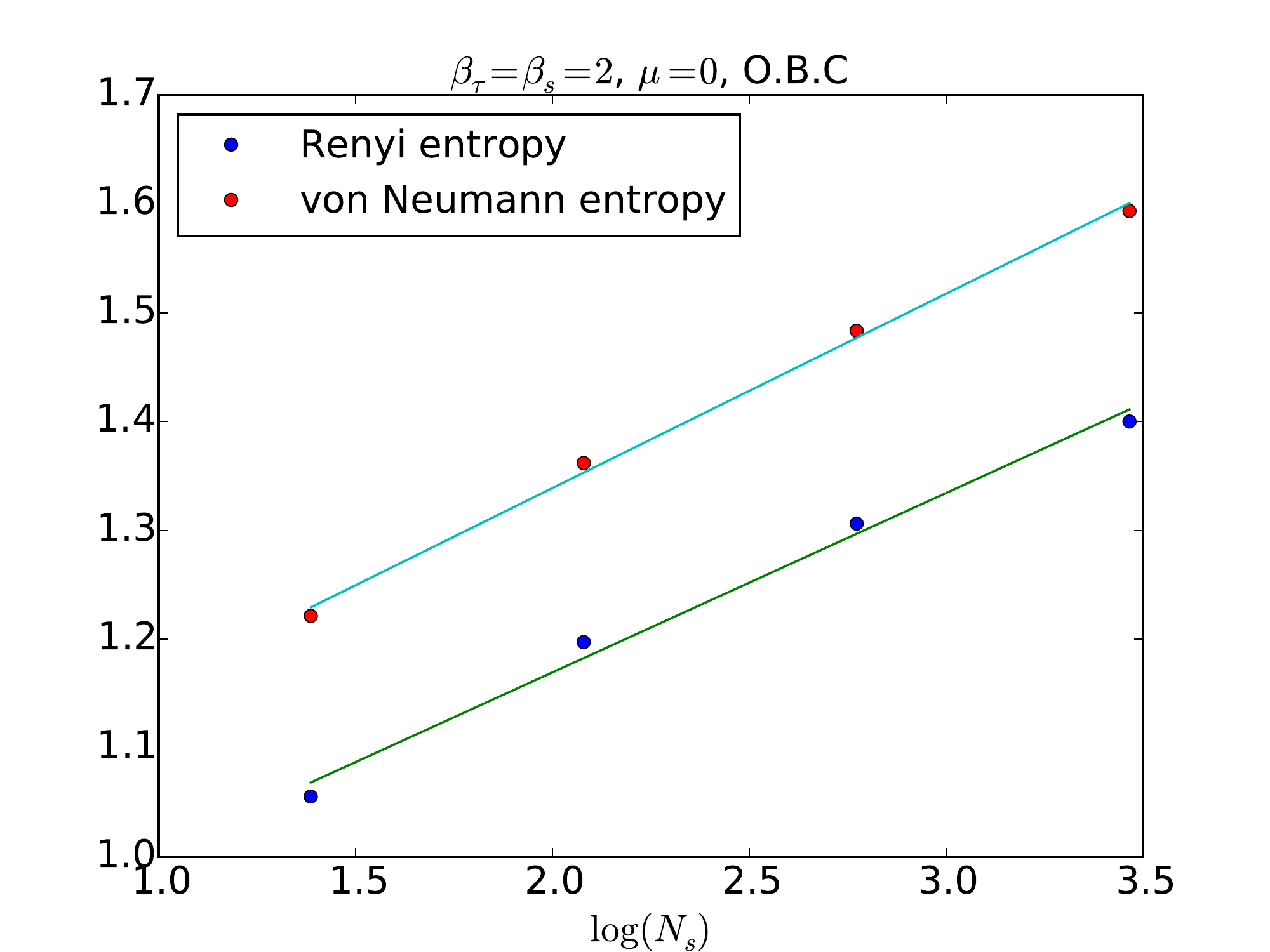}    
	\includegraphics[width=0.4\textwidth]{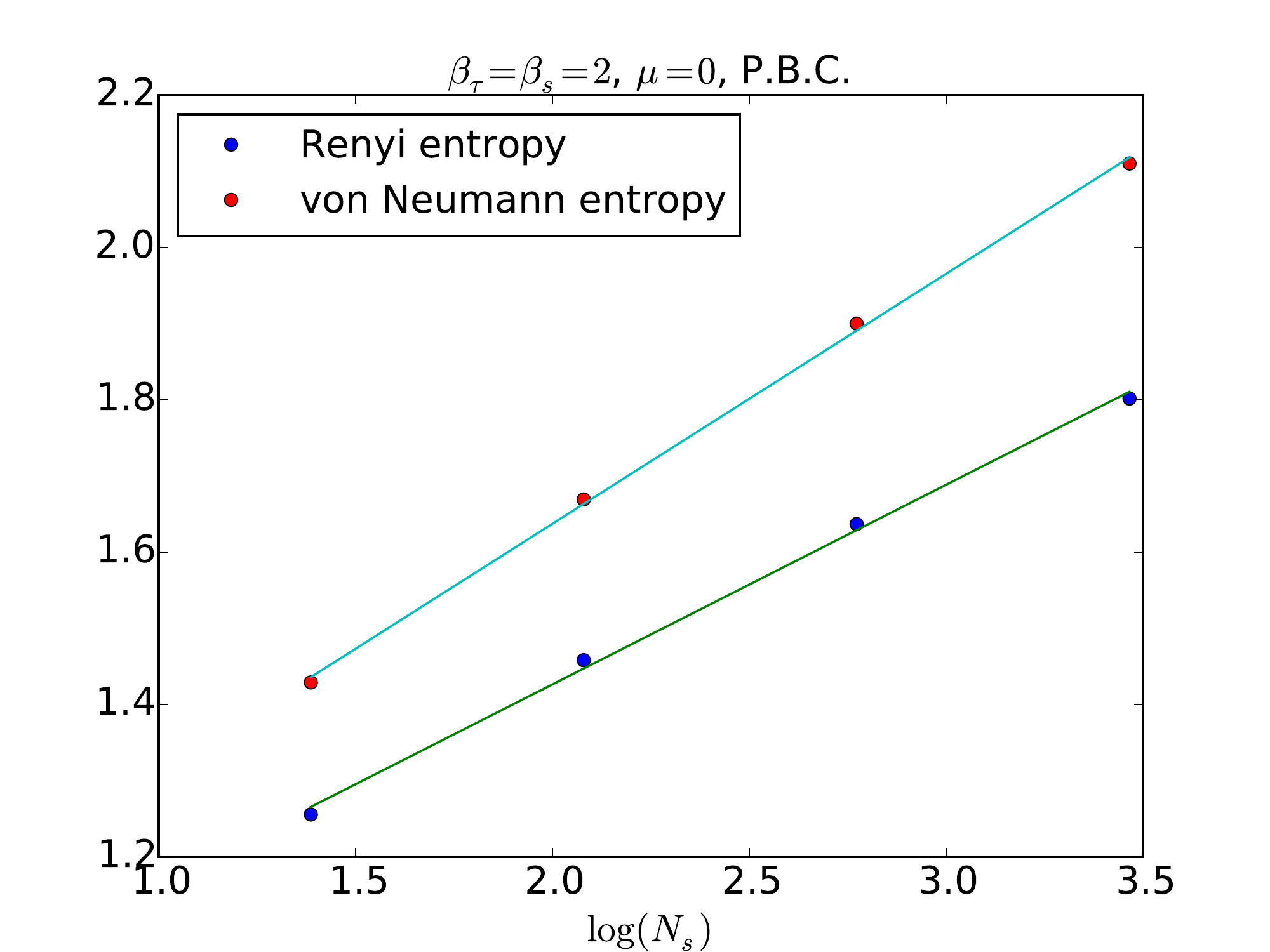}
	\caption{\label{fig:renyi12} The Von Neumann and second order Renyi EE for OBC (left) and PBC (right). }
	\end{center}
	\end{figure}

\begin{figure}[h]
\vskip-10pt
\begin{center}
\includegraphics[width=0.4\textwidth]{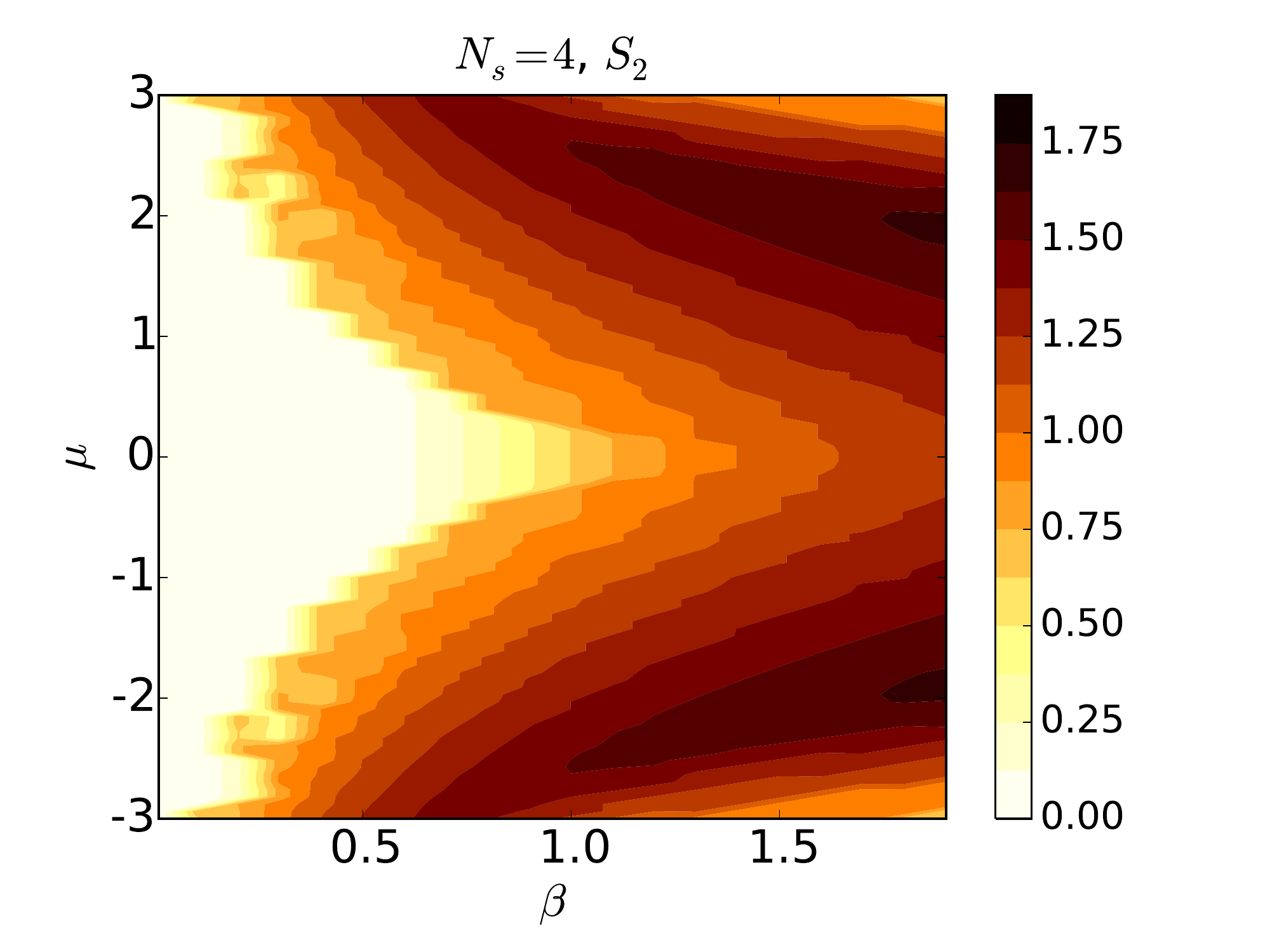}    
\caption{\label{fig:renyi3} The second order R\'{e}nyi EE ($S_2$) with $N_s=4$ and OBC  in the $\mu-\beta$ plane.}
\end{center}
\end{figure}

The TRG can be used to calculate various types of entanglement entropy (EE). 
We consider the subdivision of  a time slice $AB$ into two parts $A$ and $B$  (two halves in our calculation). 
After tracing over $B$, we obtain the reduced density matrix $\hat{\rho}_A\equiv {\rm Tr}_B \hat{\rho}_{AB}$
and we can calculate the  Von Neumann entropy
\begin{equation}
S_{ von Neumann}=-\sum_i \rho_{A_i }\ln(\rho_{A_i}), \label{Eq:entropy}
\end{equation}
where $\rho_{A_i }$ are the eigenvalues of $\hat{\rho}_A$. 
We then use the TRG blocking method until $A$ and $B$ are each reduced to a single site.
With this method, we can only calculate the entanglement entropy for regions that contain 
a number of sites that is a power of 2. 
We can also compute the $n$ -th order R\'enyi EE  defined as:
\begin{equation}
S_n(A)\equiv \frac{1}{1-n}\ln(\Tr(\hat{\rho}_A^n)) \ . 
\end{equation}
In the limit where ${n\rightarrow 1^+}$, we recover the von Neumann EE. 
Numerical results for the $O(2)$ model in 1+1 dimensions with a chemical potential are shown in Fig.  \ref{fig:renyi12} 
for open (OBC) and periodic boundary conditions  (PBC) in the Kosterlitz-Thouless (KT) phase. 
The approximately linear behavior in $\ln(N_s)$ is consistent with the Calabrese-Cardy scaling which predicts
\begin{equation}
S_n(N_s) = 
\begin{cases}
	K_n+\frac{c(n+1)}{6n}\ln(N_s) & \text{for PBC} \\
    K_n'+\frac{c(n+1)}{12n}\ln(N_s) & \text{for OBC}.
    \label{eq:lcft}
\end{cases}
\end{equation}
The constants $K$ are non-universal 
and different in the four situations considered ($n$=1, 2 with PBC and OBC). 
$S_2$ can also be used to approximately localize the superfluid phase (where the particle density increases with the chemical potential). This is illustrated in Fig. \ref{fig:renyi3}.  It shares some features with the $O(3)$ model with a chemical potential \cite{falko3} which can also be quantum simulated \cite{laflamme}. 

In order to check  with the Density Matrix Renormalization Group (DMRG) method, 
the time continuum limit can be achieved by increasing $\beta_\tau$ while keeping constant the products $\beta_s \beta_\tau=2\tilde{J}/\tilde{U}$ and $\mu \beta_\tau =\tilde{\mu}/\tilde{U}$. This defines the rotor Hamiltonian:
\begin{equation}
	\hat{H}=\frac{\tilde{U}}{2}\sum_x \hat{L}_x^2-\tilde{\mu}\sum_x 		\hat{L}_x-2\tilde{J}\sum_{\left<xy\right>}\cos(\hat{\theta}_x-			\hat{\theta}_y) \ ,
\label{eq:rotor}
\end{equation}
with $[\hat{L}_x,{\rm e}^{i\hat{\theta} _y}]=\delta_{xy}{\rm e}^{i\hat{\theta} _y}$.  For quantum simulation purposes, these commutation relations can be approximated 
for finite integer spin \cite{PhysRevA.90.063603}.
In the following we focus on the spin-1 approximation which can also be implemented 
in the classical system by setting the tensor elements to zero for space and time indices strictly larger than 1 in absolute value. The correspondence between the  two methods 
can be checked with a Density Matrix Renormalization Group (DMRG) method which 
optimizes the EE and allows one to calculate observables for any number of sites. When applied to $S_2$ with OBC, large subleading corrections 
become apparent  \cite{preprint,draft} for the intermediate values of $L$ not shown in Fig. \ref{fig:renyi3}. 
\section{The Polyakov loop in the Abelian Higgs model}
The TRG method was used to calculate the partition function of the Abelian Higgs model \cite{PhysRevD.92.076003}.
We attach a $B^{(\Box)}$ tensor  to every plaquette,
 a $A^{(s)}$ tensor to the spatial links (they cross these links and point in the Euclidean time direction)
 and a $ A^{(\tau)}$ tensor to the temporal links (they  point in the spatial direction, see figures in \cite{PhysRevD.92.076003}). 
The partition function can be written as
\vskip-10pt
        \begin{eqnarray}
         &&   Z= ( I_{0}(\beta_{pl})I_0(2\kappa_s)I_0(2\kappa_\tau))^V \times \nonumber \\
          &&  {\rm Tr}\left[ \prod_{h,v,\Box}A^{(s)} _{m_{up}m_{down}} A^{(\tau)} _{m_{right}m_{left}}B^{(\Box)}_{m_1m_2m_3m_4}\right].
        \end{eqnarray}
A few remarks are in order. 
The quantum numbers on the links are completely determined by the quantum numbers on the plaquettes. 
The plaquette quantum numbers are the dual variables. The reformulation is manifestly gauge invariant. If we impose periodic boundary conditions on the plaquettes, we can only have neutral states (Gauss law). For related questions in QED, see  Ref. \cite{cstar}. 
It is however possible to probe the charge sectors by introducing Polyakov loops.

\begin{figure}
 \begin{center}
\includegraphics[width=1.3in]{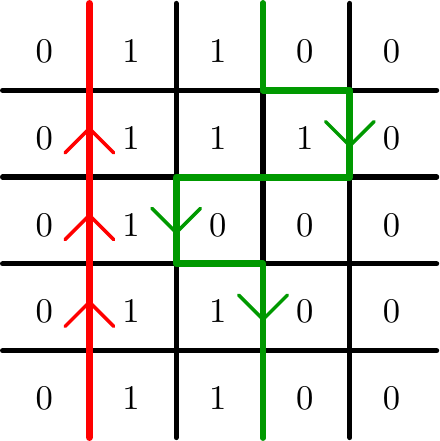}
\hskip19pt
  \includegraphics[width=1.3in]{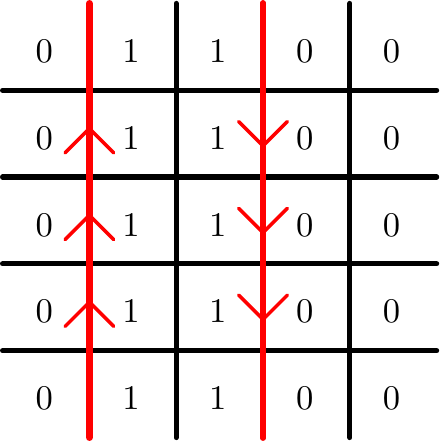}
  \caption{\label{fig:ploop}Polyakov  loop (red), current (green) in the opposite direction (left); compensation  by another Polyakov loop (right).}
  \end{center}
  \end{figure}

We consider Polyakov loops wrapping around the Euclidean time direction: $\langle P_i \rangle = \langle \prod_{j} U_{(i,j),\tau} \rangle$.  
With spatial periodic boundary conditions, the insertion of the Polyakov  loop shown in red in  Fig. \ref{fig:ploop} forces the presence of a scalar current (green) in the opposite direction (left) or it may be compensated  by another Polyakov loop in the opposite direction (right).

\begin{figure}[h]
\begin{center}
\includegraphics[width=2.5in]{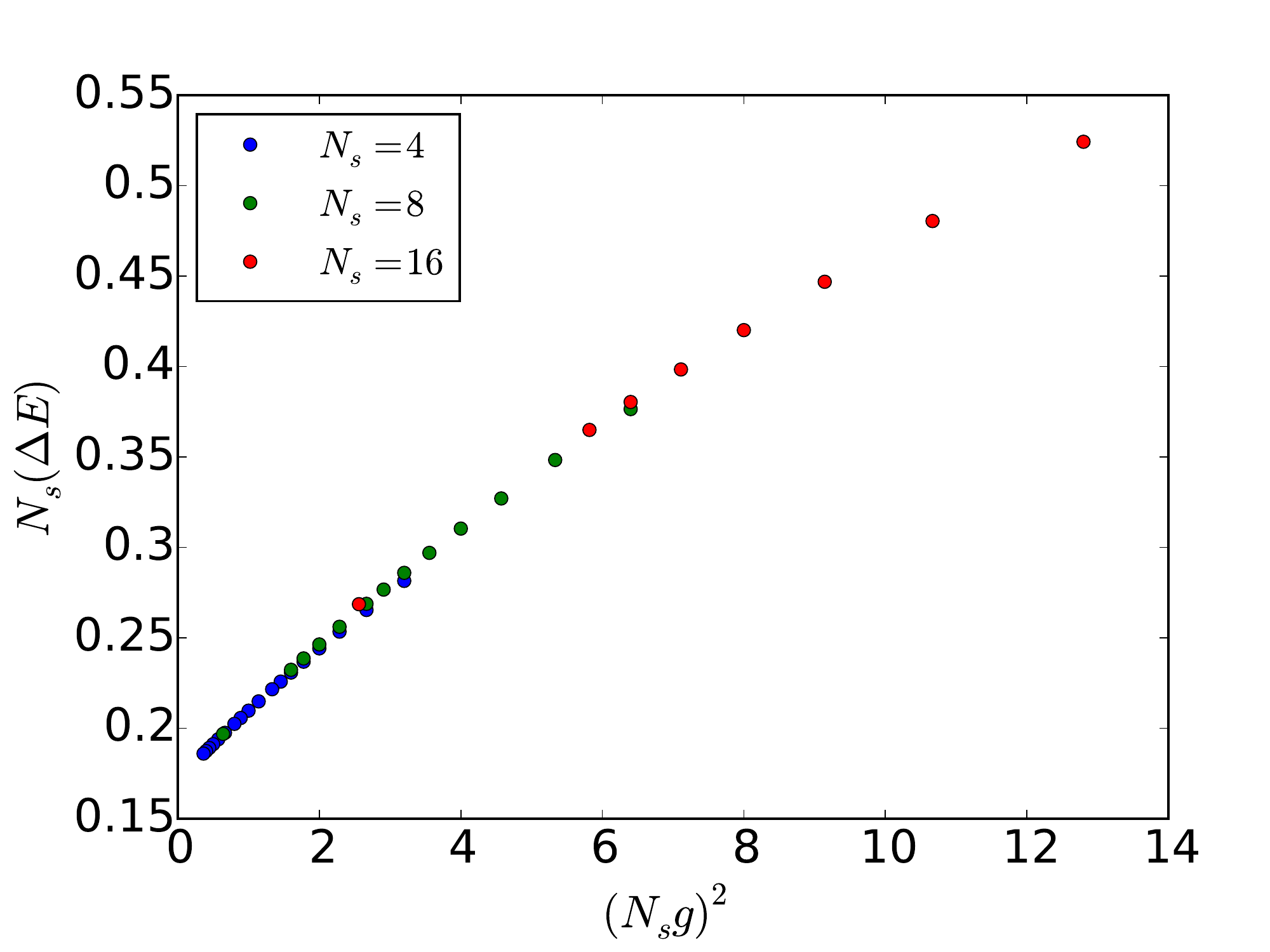}
\includegraphics[width=2.5in]{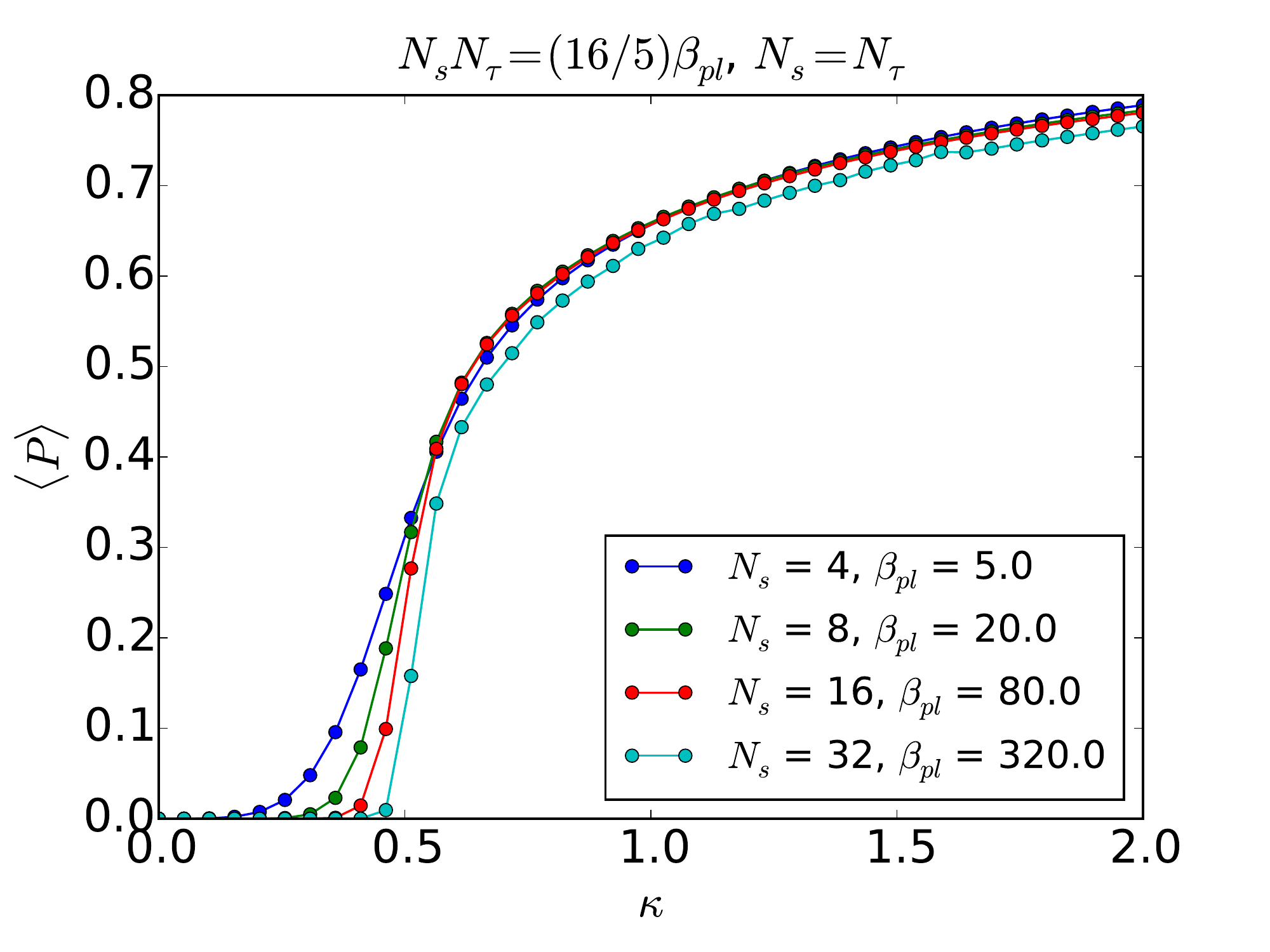}
\caption{(left) The Polyakov loop as function of $\kappa=\beta_s/2$, the increase in sharpness with volume makes it look like an order parameter; (right) data collapse for $N_{s} \Delta E$. }
\label{fig:l2}
\end{center}
\end{figure}

Some interesting data collapse  has been found for the Polyakov loop $P$. The relation  with the mass gap $\Delta E$ is 
 that $-\ln (P) \simeq C+N_\tau(\Delta E)$. We then use the fact that in absence of gauge coupling, the gap in the KT phase decreases like $1/N_s$, while the gauge fields create horizontal lines of 1's in Fig. 3 and so $ \Delta E\simeq A/N_s +B g^2 N_s+ ...$ and we obtain a 
data collapse for  $N_s \Delta E =F(g^2 N_s^2) $. Numerical calculations shown in Fig. \ref{fig:l2} give support to this idea.

\section{Quantum Simulators}

There has been a recent interest in using cold atoms trapped in optical lattices for quantum simulating spin and gauge models  studied by lattice gauge theorists \cite{Tagliacozzo:2012vg}. 
This means using the interaction of polarizable cold atoms trapped in a periodic potential and their tunneling properties to built a system evolving at real time with a Hamiltonian similar to one of the models considered. 
There are no sign problems
and real time evolution occurs at physical time. 
So far  linear sizes reached in experiment can be of order 100-200 and are expected to reach 1000 soon. 
Our approach is based on the TRG formulation of lattice gauge theory and is manifestly gauge invariant.

So far, the remarkable theory/experiment reached for the Bose-Hubbard model is just a source of inspiration in the context of lattice gauge theory and a proof of principle is needed. We hope that Ref. \cite{preprint} is a step in this direction. 
In this recent preprint, we compared the second-order R\'enyi EE, $S_2$, of the classical $O(2)$ model with a chemical potential on a  1+1 dimensional lattice, and a quantum Bose-Hubbard Hamiltonian that can be simulated with cold atoms on a one-dimensional optical lattice. Both models have a superfluid phase where we expect $S_2$ to follow the Calabrese-Cardy scaling.  Near half-filling and for a small hopping parameter $J$, $S_2$ is almost identical for both models. 
We proposed to amend the existing experimental setup to measure $S_2$ by adiabatically reaching the ground state of twin tubes {\it half-filled} with $^{87}$Rb atoms with small $J$; this is in contrast to existing experiments with density one at larger $J$. 

\section{Conclusions}
The TRG formulation  allows reliable calculations of the phase diagram and spectrum of the 1+1 dimensional $O(2)$ model with a chemical potential. Calculations of the von Neumann and R\'enyi EE for the $O(2)$ model in the superfluid phase at increasing $N_s$ seem consistent with CFT of central charge 1. 
We have proposed a gauge-invariant approach for the quantum simulation of the abelian Higgs model.
Calculations of the Polyakov loop at finite $N_s$ and small gauge coupling shows  interesting behavior.  
We obtained a nice data collapse at weak gauge coupling. 

\vskip5pt
{\it Acknowledgments}. This research was supported in part  by the Department of Energy
under Award Numbers DOE grant DE-FG02-05ER41368, DE-SC0010114 and DE-FG02-91ER40664, the NSF under grant DMR-1411345 and by the Army Research Office of the Department of Defense under Award Number W911NF-13-1-0119.  
L.-P. Yang was supported by Natural Science Foundation for young scientists of China (Grants No.11304404) and Research Fund for the Central Universities(No. CQDXWL-2012-Z005). Parts of the  numerical calculations were done  at the Argonne  Leadership  Computational Facilities.

\providecommand{\href}[2]{#2}\begingroup\raggedright\endgroup

\end{document}